\documentclass[useAMS,usenatbib]{mn2e}
\usepackage[dvips]{graphicx}
\usepackage{amsmath}
\usepackage{amsfonts}
\usepackage{amssymb}

\title[Bining is Sinning]{Binning is sinning: morphological light-curve 
distortions due to finite integration time}
\author[David M. Kipping]{David M. Kipping$^{1,2}$\thanks{E-mail:
dkipping@cfa.harvard.edu} \\
$^{1}$Harvard-Smithsonian Center for Astrophysics, 
       60, Garden Street, Cambridge, MA 02138, USA \\
$^{2}$Department of Physics and Astronomy, University College London,
       Gower Street, London WC1E 6BT, UK}
\begin{document}

\date{Accepted 2010 June 22. Received 2010 June 21; in original form 2010 June 21}

\pagerange{\pageref{firstpage}--\pageref{lastpage}} \pubyear{2010}

\maketitle

\label{firstpage}

\begin{abstract}

We explore how finite integration times or equivalently temporal binning induces 
morphological distortions to the transit light-curve. These distortions, if 
uncorrected for, lead to the retrieval of erroneous system parameters and may 
even lead to some planetary candidates being rejected as ostensibly unphysical. 
We provide analytic expressions for estimating the disturbance to the various 
light-curve parameters as a function of the integration time. These effects are 
particularly crucial in light of the long-cadence photometry often used for 
discovering new exoplanets by, for example, \emph{Convection Rotation and 
Planetary Transits (COROT)} and the \emph{Kepler Mission} (8.5 and 30\,min). 
One of the dominant effects of long integration times is a systematic 
underestimation of the light-curve-derived stellar density, which has 
significant ramifications for transit surveys. We present a discussion of 
numerical integration techniques to compensate for the effects and produce 
expressions to quickly estimate the errors of such techniques, as a function of 
integration time and numerical resolution. This allows for an economic choice of 
resolution before attempting fits of long-cadence light-curves. We provide a 
comparison of the short- and long-cadence light-curves of TrES-2b and show that 
the retrieved transit parameters are consistent using the techniques discussed 
here.

\end{abstract}


\section{Introduction}

Transiting extrasolar planets have become a powerful tool since the first 
discovery by \citet{cha00} and \citet{hen00}. The detection of a transit allows 
for a measurement of the planet-star ratio-of-radii, which may be used to infer 
the planetary radius once the stellar radius has been determined. 
Multi-wavelength measurements may be used to retrieve a spectrum of the light 
being absorbed by the planetary atmosphere and thus a deduction of the molecular 
constituents (e.g. \citealt{tin07}). Timing of the eclipse minima and duration 
permits for the detection of other bodies in the system \citep{ago05,hol05} and 
companion moons \citep{kip09a,kip09b}. In recent years, various authors have 
published increasingly ingenious methods of characterizing exoplanets through 
transits, including thermal emission detections \citep{dem06}, spin-orbit 
alignment \citep{win05} oblateness measurements \citep{sea02,car10} and many 
more.

As we look for these small whispers of the exoplanet's nature, we are in fact 
examining the morphology of the transit light-curve in increasing detail. It is 
therefore paramount that the methods for modeling these light-curves are in the 
most precise form possible.

In this paper, we explore the effects of finite integration time on the transit 
light-curve, or equivalently the act of temporal binning. Long-cadence (LC) 
data smears out the transit light-curve signal into a broader shape which will 
lead to an erroneous retrieval of the system parameters, if unaccounted for.

We will first outline the nature of the morphological distortions in 
\S\ref{sec:sec2} and then derive analytic expressions for the approximate 
systematic error in the retrieved parameters, as a function of integration time.  
It is shown that the effect can be significant enough to even lead to some 
planetary candidates being rejected as being unphysical, unless the effect is 
correctly accounted for. Obviously, this is critical for missions like 
\emph{CoRoT} and \emph{Kepler} which employ LC photometry to detect new 
transiting planets. In \S\ref{sec:modeling}, we will discuss numerical 
techniques to correctly model the transit light-curve and derive the errors of 
these methods as a function of integration time and numerical resolution. 
These expressions allow for a simple evaluation of the expected consequences of 
finite integration time and thus will guide observers in choosing how to deal 
with this issue.

\section{The Effects of Finite Integration Time}
\label{sec:sec2}

\subsection{Ingress/Egress Durations}

For a transit light-curve, there are four critical contact points which define 
the overall shape, which represent the points where the time derivative is 
discontinuous.  Physically speaking, contact points I and IV occur when the 
sky-projected planet-star separation is equal to the stellar radius plus to 
planetary radius.  Contact points II and III occur when this parameter equals 
the stellar radius minus the planetary radius. Defining $W$ as the average of 
the durations between the 1$^{\mathrm{st}}$-to-4$^{\mathrm{th}}$ and 
2$^{\mathrm{nd}}$-to-3$^{\mathrm{rd}}$ contacts \citep{kip10}, $t_c$ as the 
mid-transit time and $\tau$ as the ingress/egress duration, we have:

\begin{align}
t_{I} &= t_c - W/2 - \tau/2 \\
t_{II} &= t_c - W/2 + \tau/2 \\
t_{III} &= t_c + W/2 - \tau/2 \\
t_{IV} &= t_c + W/2 + \tau/2
\end{align}

\begin{figure*}
\begin{center}
\includegraphics[width=16.8 cm]{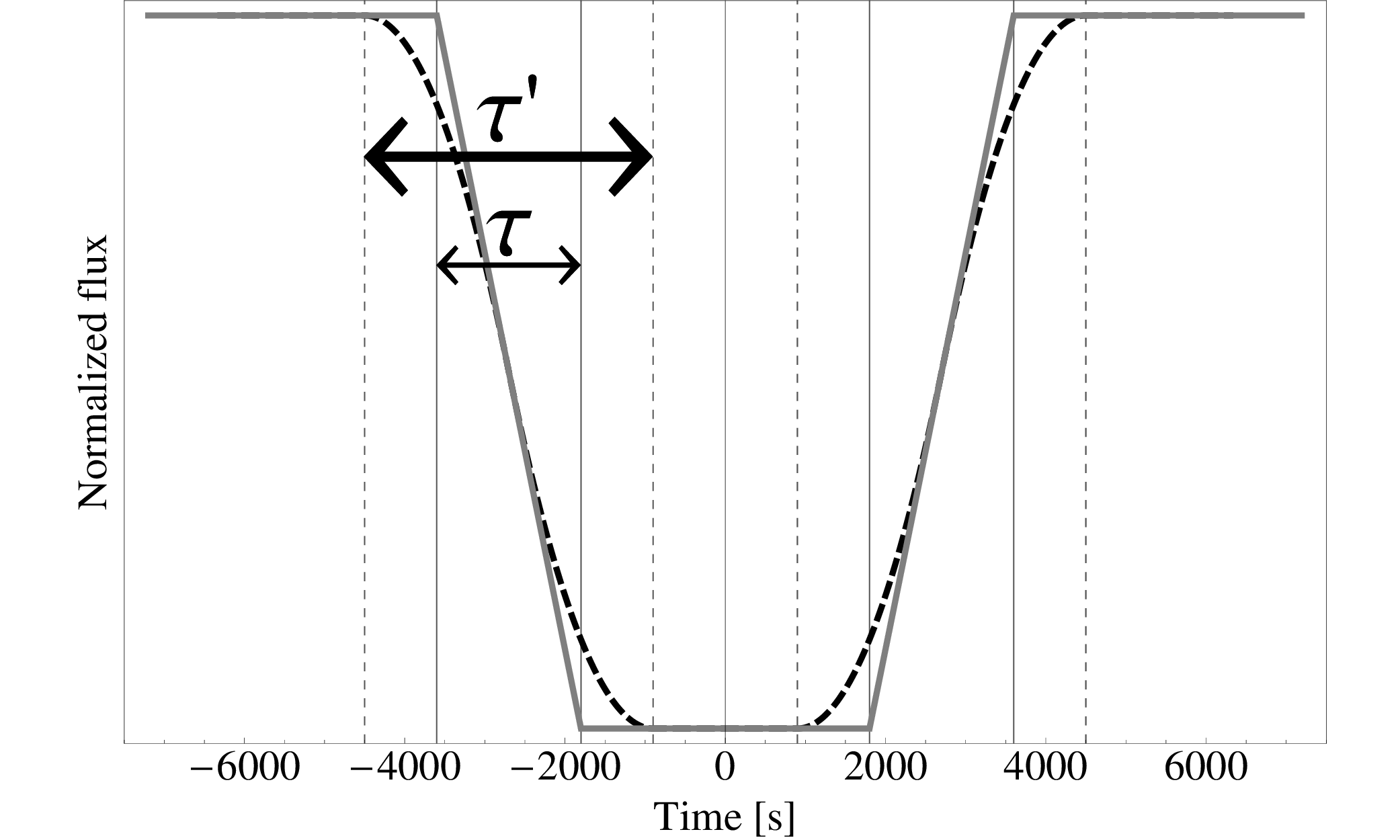}
\caption{\emph{A trapezoid approximated light-curve with a one hour flat-bottom 
duration and 30 minute ingress/egress duration, $\tau$, is shown in solid. The 
dashed line shows the light-curve morphology for an integration time of 
30\,minutes.  The apparent ingress/egress duration, $\tau'$, can be seen to have 
doubled purely as a consequence of the integration time.}}
\label{fig:fig1}
\end{center}
\end{figure*}

The principal effect of finite integration time is to smear out the light-curve 
into a broader shape (see Figure~\ref{fig:fig1}).  The apparent ingress and 
egress durations will increase and additional curvature will be introduced into 
the light-curve wings. The ingress/egress stretching can be considered in terms 
of the apparent positions of the contact points being temporally shifted from 
their true value. The magnitude of this time shift is dependent on the relative 
phase difference between the sampling and the transit signal. If we assume that 
a large number of transits observed with LC photometry are folded about the 
orbital period, as is typical in transit detection, then the effect becomes much 
more predictable with the deviation averaging out to $\mathcal{I}/2$.

Under these conditions, contact points I and IV will appear to move outwards 
from $t_c$ by one half of the integration time each, $\mathcal{I}/2$. 
Conversely, contact points II and III will appear to move inwards by the same 
amount.  Let us define the apparent contact points as $t'$:

\begin{align}
t_{I}' = t_{I} - \mathcal{I}/2 \\
t_{II}' = t_{II} + \mathcal{I}/2 \\
t_{III}' = t_{III} - \mathcal{I}/2 \\
t_{IV}' = t_{IV} + \mathcal{I}/2
\end{align}

\citet{sea03} showed that several physical system parameters may be retrieved by 
analysis of the light-curve morphology.  Specifically, they define $t_T = t_{IV} 
- t_{I}$ and $t_F = t_{III} - t_{II}$.  Let us define the apparent observed 
values as $t_T'$ and $t_F'$ respectively. \citet{sea03} showed that the 
semi-major axis of the planetary orbit divided by the stellar radius, $a_R$, the 
transit impact parameter, $b$, and the ratio-of-radii, $p$, are all determined 
by manipulation of $t_T$, $t_F$ and transit depth, $\delta$.  These parameters 
may then be used to determine orbital inclination, $i$, and stellar density, 
$\rho_*$.  Recently, \citet{kip10} extended these equations to account for 
orbital eccentricity and we will here employ these more general expressions in 
our analysis:

\begin{align}
b^2 &= \frac{(1-p)^2 - \frac{\sin^2[(t_F \pi \sqrt{1-e^2})/(P \varrho_c^2)]}{\sin^2[(t_T \pi \sqrt{1-e^2})/(P \varrho_c^2)]} (1+p)^2}{ 1-\frac{\sin^2[(t_F \pi \sqrt{1-e^2})/(P \varrho_c^2)]}{\sin^2[(t_T \pi \sqrt{1-e^2})/(P \varrho_c^2)]} } \\
a_R^2 &= \frac{(1+p^2) - b^2}{\varrho_c^2 \sin^2[(t_T \pi \sqrt{1-e^2})/(P \varrho_c^2)]} + \frac{b^2}{\varrho_c^2} \\
\rho_{*} &= \frac{3 \pi}{G P^2} a_R^3 - p^3 \rho_P
\end{align}

We here derive what these retrieved parameters would be if the effect of finite 
integration time was ignored.  In the above expressions, we replace $t_T$ \& 
$t_F$ with $t_T' = (t_T + \mathcal{I})$ \& $t_F' = (t_F - \mathcal{I})$. The 
new expressions then have the remaining $t_T$ \& $t_F$ terms written out in 
terms of the true values of $b$, $a_R$, etc.  This leaves us with 
$b'(b,a_R,...)$, $a_R'(b,a_R,...)$, etc:

\begin{align}
b'^{2} &= \frac{\Bigg[(1+p)^2 \frac{\sin^2\Big[\frac{\sqrt{1-e^2} \mathcal{I} \pi}{P \varrho_c^2} - \arcsin\Big(\frac{\sqrt{(1-p)^2 - b^2}}{a_R \varrho_c \sin i}\Big)\Big]}{\sin^2\Big[\frac{\sqrt{1-e^2} \mathcal{I} \pi}{P \varrho_c^2} + \arcsin\Big(\frac{\sqrt{(1+p)^2 - b^2}}{a_R \varrho_c \sin i}\Big)\Big]}- (1-p)^2\Bigg]}{\Bigg[\frac{\sin^2\Big[\frac{\sqrt{1-e^2} \mathcal{I} \pi}{P \varrho_c^2} - \arcsin\Big(\frac{\sqrt{(1-p)^2 - b^2}}{a_R \varrho_c \sin i}\Big)\Big]}{\sin^2\Big[\frac{\sqrt{1-e^2} \mathcal{I} \pi}{P \varrho_c^2} + \arcsin\Big(\frac{\sqrt{(1+p)^2 - b^2}}{a_R \varrho_c \sin i}\Big)\Big]}- 1\Bigg]} \\
a_R'^2 &= \frac{\frac{\Bigg[(1+p)^2 \sin^2\Bigg(\frac{\sqrt{1-e^2} \pi \mathcal{I}}{\varrho_c^2 P} - \arcsin\Big(\frac{\sqrt{(1-p)^2 - b^2}}{a_R \varrho_c \sin i}\Big)\Bigg)-4p\Bigg]}{\sin^2\Bigg(\frac{\sqrt{1-e^2}\pi \mathcal{I}}{\varrho_c^2 P} + \arcsin\Big(\frac{\sqrt{(1+p)^2-b^2}}{a_R \varrho_c \sin i}\Big)\Bigg)}-(1-p)^2}{\varrho_c^2 \frac{\sin^2\Bigg(\frac{\sqrt{1-e^2}\pi \mathcal{I}}{\varrho_c^2 P} - \arcsin\Big(\frac{\sqrt{(1-p)^2 - b^2}}{a_R \varrho_c \sin i}\Big)\Bigg)}{\sin^2\Bigg(\frac{\sqrt{1-e^2}\pi \mathcal{I}}{\varrho_c^2 P} + \arcsin\Big(\frac{\sqrt{(1+p)^2 - b^2}}{a_R \varrho_c \sin i}\Big)\Bigg)}-\varrho_c^2} \\
\rho_{*}' &\simeq \frac{3 \pi}{G P^2} a_R'^3
\end{align}

Setting $\mathcal{I} = 0$ returns the original results as expected. 
Unfortunately, these equations are somewhat overly complex for us to draw any 
physical intuition. To proceed, let's consider a typical case example by using 
the system parameters from one of the \emph{Kepler} planets, since these are 
discovered using long-cadence data. The following example is for the assumption 
of zero limb darkening, which is a very poor one for the \emph{Kepler} bandpass.  
The effects of limb darkening will be discussed later.

\begin{figure*}
\begin{center}
\includegraphics[width=16.8 cm]{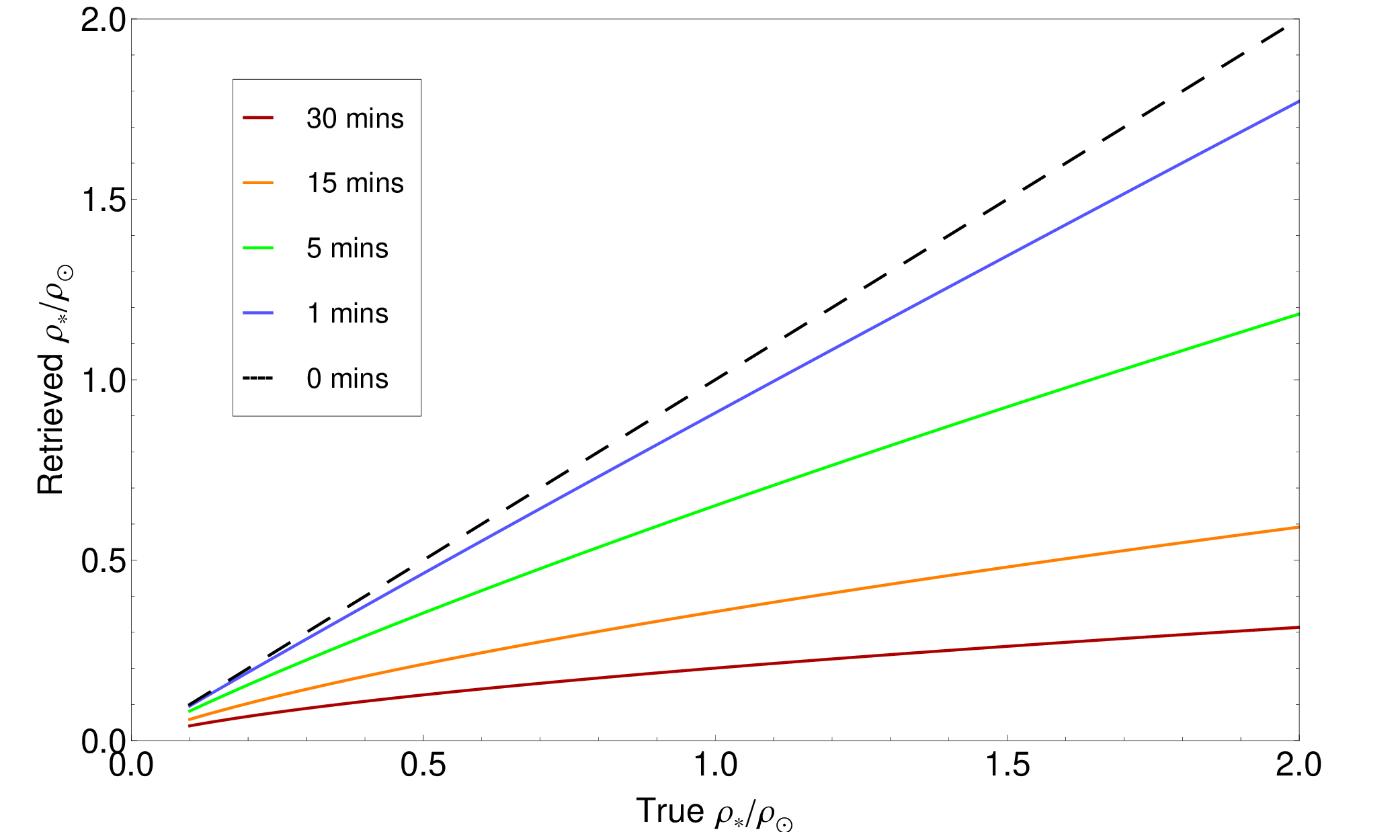}
\caption{\emph{As an example, we use the system parameters of Kepler-5b to show 
the effect on the retrieved stellar density as a result of long integration 
times, in the case of no limb darkening.  The 1 min line appears to produce 
results within the typical uncertainties of the derived stellar density.}} 
\label{fig:fig2}
\end{center}
\end{figure*}

In Figure~\ref{fig:fig2}, we plot the retrieved stellar density, $\rho_*'$, as a 
function of the true stellar density, $\rho_*$. We fix all other parameters to 
be that of Kepler-5b, as reported by \citet{koc10}. The effect can be seen to be 
highly significant, causing the retrieved stellar density to be underestimated 
by a factor which borders on an order-of-magnitude. This scale of 
underestimation is sufficient to completely reject some planetary candidates as 
unphysical. However, we note that in reality the underestimation of $\rho_*$ 
will not be this severe due to countering effects of limb darkening suppression 
discussed later in \S\ref{sub:LD}.

Looking at Figure~\ref{fig:fig2} again, let us explore the physical reasons for 
this underestimation. We begin our line of thought by considering the impact 
parameter, $b$. If the ingress/egress duration is elongated, what do we expect 
to happen to the derived impact parameter? Consider two extreme cases. When a 
planet transits with a very low impact parameter, we have essentially an 
equatorial transit. This means the vector describing the sky-projected 
planetary velocity is nearly perpendicular to the stellar limb. As a result, 
the planet crosses the limb expediently. In contrast, if we have a near-grazing 
transit, the mutual angle between the stellar limb and the sky-projected 
planetary velocity vector has become more acute, which has the effect of making 
the limb-crossing time much longer. Therefore, stretching the ingress/egress 
duration causes $b' > b$.

Having established this point, consider the effect on $a_R$.  The simplest way 
to understand the effect on this parameter is to appreciate that $b$ and $a_R$ 
exhibit an extremely strong negative correlation, as demonstrated by 
\citet{car08}. So the act of increasing the ingress/egress duration will 
increase $b$ and therefore decrease $a_R$. Finally, from equation (11) we know 
that $\rho_* \propto a_R^3$.  

\subsection{Limb Darkening Effects}
\label{sub:LD}

For transit observations at visible wavelengths, limb darkening is quite 
pronounced producing a well-known curvature in the flat-bottom part of the 
light-curve. For large integration times, the curvature is smeared out, 
producing a flatter transit trough morphology.

If we were trying to fit such a transit light-curve, and the limb darkening 
parameters are exactly known, our hypothetical fitting algorithm would move 
towards a case where limb-darkening is least effective. This occurs for 
near-equatorial transits. This is because as the transit impact parameter 
becomes larger and larger, we move closer and closer to the limb of the star 
where limb darkening accentuates. Therefore, LC data pushes $b'$ towards a more 
equatorial value due to limb darkening i.e. $b' < b$.

\subsection{Two Countering Effects}

In conclusion, finite integration times have two countering effects on $b'$ and 
thus $a_R'$ and $\rho_*'$ as well. The ingress and egress smearing causes the 
ingress to appear larger, which occurs for more grazing transits.  In contrast, 
the limb darkening is suppressed, which occurs for more equatorial transits.  
Whilst limb darkening is important, especially for transit surveys like 
\emph{CoRoT} and \emph{Kepler} which operate at visible wavelengths, the 
fundamental change in the transit morphology is sufficiently large that it will 
tend to dominate.  This is because the amplitude of the curvature in the 
light-curve trough is usually at least an order-of-magnitude less than the 
amplitude of the transit signal itself.  Therefore, the consequences of smearing 
the overall transit signal will tend to dominate over the suppression of limb 
darkening.

The general rule of thumb that $b'$ is overestimated will break-down when we 
have $\tau \gg \mathcal{I}$ but $t_F \sim \mathcal{O}[\mathcal{I}]$, which 
occurs for near-grazing transits.  In such a case, the fractional change to the 
ingress durations is minimized but the change in the transit trough curvature is 
maximized.

To exactly calculate the net consequence of these two effects though, the 
integration time should be included when we generate our model light-curves 
rather than attempting ad-hoc corrections post-analysis.  This appears to be the 
only way to completely account for the effect in a reliable manner.

\subsection{Consequences for the Transit Depth}

We will briefly comment on the effect of finite integration time on the transit 
depth.  Assuming no limb darkening is present and $t_F > \mathcal{I}$, the 
transit depth is completely unaffected by the long integration time.  This has 
important consequences for secondary eclipses where the light-curve is 
unaffacted by stellar limb darkening.

For cases where we have limb darkening, the net effect on the retrieved $p'$ 
will depend on whether $b' > b$ or $b < b'$. Additionally, the effect will be a 
function of what assumptions were used in the fitting algorithm (e.g. the fixing 
of various parameters), the true impact parameter and the limb darkening 
coefficients. Given the large number of correlated factors, predicting the 
effect of integration time becomes less reliable and we must fit the transit 
light-curve with a model which accounts for integration time in the first place.

\subsection{Observed Effects with \emph{Kepler}}

We point out that the effects of long integration times have already been 
observed by the \emph{Kepler Mission}. Figure~4 of \citet{gil10}, shows the 
transit light-curve of long-cadence and short-cadence data for the same planet, 
TrES-2b. The long-cadence light-curve exhibits a broader shape with the apparent 
position of the contact points shifted by $\sim \mathcal{I}/2$, as predicted by 
our model. Notice also that the curvature in the transit trough, due to limb 
darkening, has also been attenuated.

\section{Accurate Transit Lightcurve Modeling}
\label{sec:modeling}

\subsection{Analytic Integration}

The critical problem we have outlined can be simply summarized by the following: 
\emph{Don't fit an unbinned model to binned data}. The model usually used to 
generate a transit light-curve is provided by \citet{man02} (MA02), which 
includes the effects of stellar limb darkening. To generate the MA02 
light-curve, we usually have a set of time stamps forming a time vector 
$\underline{t}$. This time vector represents instantaneous moments rather than 
integrated time. The $\underline{t}$ vector is converted into a vector of 
instantaneous true anomalies, $\underline{f}$, by solving Kepler's equation 
numerically. $\underline{f}$ is converted to a $\underline{z}$ array, where 
$z(f)$ is the instantaneous sky-projected planet-star separation. Finally, the 
MA02 equations provide us with $\underline{F}$, where $F$ is the instantaneous 
flux.  We summarize the sequence of events as:

\begin{equation}
\underline{t} \rightarrow \underline{f} \rightarrow \underline{z} \rightarrow 
\underline{F}(\underline{t}) \nonumber
\end{equation}

Now that we have established the mechanism of generating of a transit 
light-curve for instantaneous time stamps, $\underline{F}(\underline{t})$, let 
us consider what the transit light-curve for integrated time stamps would be, 
$\underline{\tilde{F}}(\underline{\tilde{t}})$. In this case, the integrated 
flux would be given by:

\begin{equation}
\underline{\tilde{F}}(\underline{\tilde{t}}) = \frac{\int_{t=\tilde{t}-\mathcal{I}/2}^{\tilde{t}+\mathcal{I}/2} F(t) \mathrm{d}t}{\int_{t=\tilde{t}-\mathcal{I}/2}^{\tilde{t}+\mathcal{I}/2} \mathrm{d}t}
\end{equation}

This equation suffers from the problem that $\underline{F}$ cannot be written as 
a function of $t$ analytically, since such a solution would require a 
closed-form solution to Kepler's equation, which is transcendental. Evaluating 
this expression for $\underline{F}$ as a function of $z$ is also not possible 
since we would find the following:

\begin{equation}
\int_{t=\tilde{t}-\mathcal{I}/2}^{\tilde{t}+\mathcal{I}/2} F(t) \mathrm{d}t = \int_{z(\tilde{t}-\mathcal{I}/2)}^{z(\tilde{t}+\mathcal{I}/2)} F(t(z)) \frac{\mathrm{d}t}{\mathrm{d}z}(z) \mathrm{d}z
\end{equation}

Whilst d$z$/d$t$ may be evaluated analytically through a chain rule expansion of 
$($d$z/$d$f) \times ($d$f/$d$t)$, the resultant expression will be as a function 
of $f$, rather than $z$.  If we knew $f(z)$, then we would be able to write out 
the integrand in a closed-form, but $f(z)$ can only be found by solving a 
quartic equation, as shown by \citet{kip08} (K08).  Unfortunately, as discussed 
in \citet{kip10}, there is no currently proposed method to correspond which 
roots refer to which orbital conjunction which makes a closed-form expression 
elusive.

The only remaining hope for a simple analytic expression would be to express the 
integral in terms of true or eccentric anomaly, which are inter-changeable. 
This would yield the following integral:

\begin{equation}
\int_{t=\tilde{t}-\mathcal{I}/2}^{\tilde{t}+\mathcal{I}/2} F(t) \mathrm{d}t = \int_{f(\tilde{t}-\mathcal{I}/2)}^{f(\tilde{t}+\mathcal{I}/2)} F(t(f)) \frac{\mathrm{d}t}{\mathrm{d}f}(f) \mathrm{d}f
\end{equation}

The integrand of this expression may be written out in a closed-form, by 
utilizing the solutions of K08:

\begin{equation}
\int_{t=\tilde{t}-\mathcal{I}/2}^{\tilde{t}+\mathcal{I}/2} F(t) \mathrm{d}t = \int_{f(\tilde{t}-\mathcal{I}/2)}^{f(\tilde{t}+\mathcal{I}/2)} F(z(f)) \frac{P}{2\pi} \frac{1}{\sqrt{1-e^2}} D(f) \mathrm{d}f
\end{equation}

Where $D(f)$ is the duration function defined by K08, $F(z)$ is given by MA02 
and $z(f)$ is well-known (e.g. \citealt{win10}). The integral limits do not 
possess a closed-form solution since once again we must solve Kepler's equation, 
but in principle the indefinite integral could be analytically evaluated and 
then the relevant limits applied after a subroutine provides numerical solutions 
to Kepler's equation. We believe that this strategy would be the most 
computationally efficient since we have obviated the need for any numerical 
integration.  However, we were unable to find a solution for the indefinite
integral for even a uniform-source case and will therefore focus the remainder 
of this discussion onto numerical techniques.

\subsection{Numerical Integration}

Having established the significant challenges regarding analytic integration, we 
now turn our attention to the use of numerical integration techniques. The 
functions we need to integrate over are in fact very well-behaved and 
well-approximated by compositions of polynomials and thus we anticipate even a 
low-resolution numerical integration technique should provide satisfactory 
accuracy.

In this subsection, we first consider the merit of Simpson's Rule or other 
Newton-Cotes based methods. We aim to avoid using nested quadrature methods like 
the Gauss-Kronrod or Clenshaw-Curtis, as the number of integrations required is 
large and we wish to avoid nested methods.  For the simplest case of Simpson's 
rule, we have:

\begin{equation}
\tilde{F}_i(N=3) = \frac{F(t_i-\mathcal{I}/2) + 4 F(t_i) + F(t_i+\mathcal{I}/2)}{6}
\end{equation}

Where $N$ denotes the number of calls needed to the MA02 code and essentially is 
a measure of the resolution of our numerical integration. This method may be 
extended to higher intervals by using Simpson's composite rule. Alternatively, 
we can extend to cubic, quartic, etc interpolations by using the Newton-Cotes 
formulas.  Each time we evaluate $F(t)$ requires another call to the MA02 
subroutine, and thus we wish to minimize the number of calls, but maximize the 
accuracy of the employed technique.

Simpson's composite rule works by splitting up our integration range into $2m$ 
subintervals, therefore requiring $N = 2m + 1$ calls to the MA02 code. The error 
on the composite Simpson's rule is given by:

\begin{equation}
\underset{\mathrm{Comp. Simpson}}{\sigma_{\tilde{F}}(N)} = F^{(4)}(\epsilon) \frac{\mathcal{I}}{180} \Big(\frac{\mathcal{I}}{N-1}\Big)^4
\end{equation}

Where $\epsilon$ is some number between $t_i - \mathcal{I}/2$ and 
$t_i + \mathcal{I}/2$.  In contrast, the Newton-Cotes formulas move through 
increasing orders by increasing the interpolation order.  For the $N=4$ case 
(which is the cubic interpolation scenario, known as Simpson's 3/8 rule), the 
equivalent errors between the two methods are:

\begin{align}
\underset{\mathrm{Newton-Cotes}}{\sigma_{\tilde{F}}(N=4)} &= F^{(4)}(\epsilon) \frac{3}{80} \mathcal{I}^5 \\
\underset{\mathrm{Comp. Simpson}}{\sigma_{\tilde{F}}(N=4)} &= F^{(4)}(\epsilon) \frac{1}{2880} \mathcal{I}^5
\end{align}

Thus for $N=4$, Simpson's composite rule offers greater accuracy than the 
Newton-Cotes based equation.  Moving through the higher orders in the 
Newton-Cotes family causes the error to have a functional dependence on 
$F^{(N)}$, i.e. the $N^{\mathrm{th}}$ differential of $F$.  So for $N>4$ it is 
not possible to give an exact comparison between the two methods since 
$F^{(N)}(t)$ is not known for any $N$.  Therefore, our only reliable comparison 
is for the $N=4$ case, from which we conclude the composite Simpson's rule is 
superior in terms of accuracy versus computational requirement.

\subsection{Error in Numerical Integration}

Let us now consider what value of $N$ should be used.  There are essentially 
three segments of the light-curve which exhibit curvature and thus would produce 
the maximum error in our numerical integrations, which employ linear piece-wise 
approximations.  

\begin{enumerate}
\item Curvature of the ingress/egress
\item Curvature of the limb-darkened light-curve trough
\item Discontinuities at the contact points
\end{enumerate}

The last of these is due to a discontinuous function and the former two are due 
to curvatures within continuous functions. We will treat these two different 
sources of `curvature' separately, although from the arguments made earlier, we 
expect the last of these effects to be the largest source of numerical error.

\subsubsection{Ingress/egress curvature}

The transit light-curve has a depth $\delta$ and an ingress duration $\tau$. For 
most of the ingress, the curvature is close to zero and essentially mimics a 
linear slope. However, near the contact points, the slope rapidly changes to a 
flat line of zero gradient. Therefore, near the contact points, the 
ingress/egress morphology causes large amounts of curvature. These points will 
exhibit the largest numerical errors in using a technique like Simpson's 
composite rule.

A suitable choice of resolution can be made by increasing $N$ until 
$\sigma_{\tilde{F}}|_{\mathrm{max}} \leq \sigma_{\tilde{F},\mathrm{obs}}$, i.e. 
our calculation should produce a flux which has a maximum systematic error which 
is less than the observational uncertainty. We will set our resolution to a 
point where it provides satisfactory accuracy even at the point of highest 
numerical error, i.e. within the ingress/egress near the contact points.

Another approach would be to use an adaptive composite Simpson's rule, for 
example like that proposed by McKeeman (1962). However, our preference here is 
to avoid using adaptive routines since they would require a new adaption for 
every single data point and fitting trial, which would be time consuming. The 
costs versus benefits of using such a method could warrant further investigation 
in the future.  Instead, we choose to use the adaption required for the most 
troublesome points, which we have already identified.  The required interval 
size in each element of the Simpson's composition should be decreased until we 
reach:

\begin{equation}
|S(a,\frac{a+b}{2}) + S(\frac{a+b}{2},b) - S(a,b)|/15 < \sigma_{\tilde{F},\mathrm{obs}}
\end{equation}

Where $S(\alpha,\beta)$ is Simpson's rule evaluated over the interval $\alpha$ to 
$\beta$.  In our case, the integral is over time and $a=t_I$ and $b=t_I + 
(\mathcal{I}_0/m)$, where $2m$ is the number of subintervals we split the 
integral into and $2m = N-1$ where $N$ is the required factor by which the 
number of calls to the MA02 code increases by. The reason for the subscript of 
$0$ by the $\mathcal{I}$ term will be explained shortly. Our requirement may be 
written as:

\begin{align}
& \frac{|S(t_I,t_I+\frac{\mathcal{I}_0}{2m}) + S(t_I+\frac{\mathcal{I}_0}{2m},t_I+\frac{\mathcal{I}_0}{m}) - S(t_I,t_I+\frac{\mathcal{I}_0}{m})|}{15} < \sigma_{\tilde{F},\mathrm{obs}} \\
&S(\alpha,\beta) =\frac{\beta-\alpha}{6} \Big[F(\alpha) + 4F\Big(\frac{\alpha+\beta}{2}\Big) + F(\beta)\Big]
\end{align}

In order to continue, we need to evaluate $F(t)$ in a closed-form, which cannot 
be achieved due to Kepler's equation.  However, there exists a special case 
where Kepler's equation does yield an exact closed-form solution and this occurs 
for circular orbits since $M = E = f$.  In such a case, we may write:

\begin{equation}
z(t) = a_R \sqrt{\sin^2\Big(\frac{2 \pi t}{P}\Big) + \cos^2i\cos^2\Big(\frac{2 \pi t}{P}\Big)}
\end{equation}

The ingress/egress morphology is dominated by the expressions pertaining to a 
uniform source.  Limb darkening does affect the ingress/egress curvature but 
this is much less than the amplitude of the uniform source transit signal. In 
the small-planet limit, MA02 provided the following approximation for the 
ingress/egress flux:

\begin{equation}
F(x) = 1+x\sqrt{p^2 - x^2}-p^2 \arccos\Big[\frac{x}{p}\Big]
\end{equation}

Where we have defined $z=1+x$ and it understood that $-p < x < p$ for the 
ingress/egress.  For the purposes of the evaluating the maximum error, we know 
that $x \simeq p$ and thus we may expand the cosine term into second order using 
a Taylor series.  Let us assume we have the simple case of $b=0$ which means 
that $i = \pi/2$.  We make further small-angle approximations to simplify the 
resultant expression for the error, which is justified since $2\pi t_I \ll P$. 
The other adjustment we need to account for is that so we have approximated 
$b=0$ and $e=0$.  To generalize the result, we consider that the effect of $b>0$ 
and $e>0$ is to stretch or shrink the ingress/egress duration by a factor 
$\tau/\tau_0$.  Therefore our expressions here are actually for $\mathcal{I}_0$, 
which may be written as $\mathcal{I}_0 = \mathcal{I} (\tau_0/\tau)$. We may now 
rewrite equation (24) as:

\begin{align}
\sigma_{\tilde{F},\mathrm{obs}} &> \Big|\frac{\psi^{5/2}}{108m^3} \Big[3 \Big(\sqrt{24 m p-9 \psi}-\nonumber \\
\qquad & 4 \sqrt{2 m p -\psi} - 6\sqrt{4m p-\psi} + \sqrt{8 m p-\psi}\Big)\Big]\Big|
\end{align}

Where we have used:

\begin{align}
\psi &= \frac{2 \pi a_R}{P} \frac{\tau_0}{\tau} \mathcal{I} \\
\frac{\tau_0}{\tau} &\simeq \frac{\sqrt{1-b^2} \sqrt{1-e^2}}{\varrho_c}
\end{align}

Due to the approximations made, we find that this equation is only stable for 
$m\geq 2$.  For any given data set, we simply need to solve equation (28) for 
$m$ with some sensible estimates of $p$, $b$, $e$, $\omega$, $a_R$ and $P$. As 
an example, for Kepler-5b, taking the quoted parameters from the \citet{koc10} 
paper, we find that even using $m=2$ provides an error of $0.1$ppm, which is 
well below the typical measurement uncertainty of $130$ppm.

\subsubsection{Limb-darkened trough-curvature}

Another part of the light-curve where we have significant curvature, and thus 
expect the maximum numerical integration errors, is the limb-darkened 
light-curve trough.  However, the peak-to-peak size of the changes in flux 
induced by the limb darkening are much lower than the transit signal itself 
(i.e. $\delta$); typically an order-of-magnitude. Further, the time scale over 
which these changes act is greater than that of the ingress/egress curvatures 
(i.e. $t_F \gg \tau$) except for grazing transits.  So we can see that, in 
general, the errors in our numerical integration techniques will be dominated by 
the ingress/egress curvatures rather than the limb-darkening-induced 
light-curve-trough curvatures.

\subsubsection{Contact point discontinuities}

The final source of variation in the light-curve gradient is that of the 
discontinuous change located at the contact points. Estimating the error due to 
this discontinuity is most easily estimated by assuming a trapezoid approximated 
light-curve and considering the location where maximal error is induced. The 
largest error (and in fact only error) will occur for measurements close to 
contact points, or more specifically $|t_i - t_M| < \mathcal{I}/2$ where $t_M$ 
is the time of one of the contact points.

Before the first contact point, we have a flat line at $F=1$ and after this 
point we have a linear slope with a gradient $-(\delta/\tau)$. The error in 
Simpson's composite rule will depend upon the relative phasing between the 
centre of the integration and the contact point, i.e. $(t_i - t_I)$. Generalized 
to any phase, the true integrated flux of the trapezoid approximated light-curve 
for the $i^{\mathrm{th}}$ time stamp is given by:

\begin{equation}
\tilde{F}_{\mathrm{true},i} = 1 - \frac{\delta}{\tau} \frac{(2 t_i + \mathcal{I})^2}{8 \mathcal{I}}
\end{equation}

For each value of $m=1,2,3...$ we choose to set the phase to be such that the 
difference between the true integrated flux and that from Simpson's method is 
maximized. Under such a condition, it may be shown that the maximum error is 
given by:

\begin{align}
\underset{\mathrm{Comp. Simp.}}{\sigma_{\tilde{F}}} &= \frac{\delta}{\tau} \frac{\mathcal{I}}{24 m^2} \\
\qquad&= \frac{\delta}{\tau} \frac{\mathcal{I}}{6 (N-1)^2}
\end{align}

For the system parameters of Kepler-5b, we find that using $m=1,2,3$ induces a 
maximal error of $371$ppm, $93$ppm and $41$ppm respectively. Given that the 
measurement uncertainties are $130$ppm \citep{koc10}, a suitable choice for 
the resolution would be $m=2$ since this means the maximum possible error of a 
data point in the least-favourable phasing would be below the measurement error.

It is interesting to see that for $m=2$ the error was $0.1$ppm for the 
ingress/egress curvature of the same system, suggesting the discontinuity error 
dominates the error budget.  Actually, this is expected from the arguments made 
earlier in this paper. Therefore, in most applications, a selection for $m$ 
based on the error induced by the contact point discontinuities will provide a 
robust integration resolution.

\subsection{Resampling}
\label{sub:resampling}

An additional method for numerically integrating the light-curve is discussed 
here. Let us consider that we have observations with integrated time stamps 
given by the vector $\underline{\tilde{t}}$. A second way of calculating 
$\underline{\tilde{F}}(\underline{\tilde{t}})$ is to resample the time vector 
into a very fine cadence, at which point we may assume $\tilde{F}=F$.  Let us 
define our temporary resampled time vector as $\underline{\tilde{t}'}$.  As an 
example, for the \emph{Kepler} data, we may choose to resample the 30 minute 
integrations into 1 minute integrations would be done by expanding each time 
stamp, $\tilde{t}_i$ into a sub-vector of 30 equally spaced time stamps with a 
mean value given by $\tilde{t}_i$.  Our new temporary time array is used to 
generate a light-curve using the normal MA02 expressions giving us 
$\underline{F'}(\underline{\tilde{t}'})$ (note that $F$ here has no tilde 
because the MA02 equations can only generate instantaneous flux, not integrated 
flux).  We then rebin the model light-curve back to the original cadence to give 
$\underline{F}(\underline{\tilde{t}})$.  Finally, we make the assumption 
$\underline{\tilde{F}}(\underline{\tilde{t}'}) \simeq 
\underline{F}(\underline{t})$, i.e. the high cadence resampled time vector 
yields a light-curve model consistent with a time vector of infinite cadence.

\begin{equation}
\underline{\tilde{t}} \underset{\mathrm{resample}}{\rightarrow} \underline{\tilde{t}'} \underset{\mathrm{MA02}}{\rightarrow} \underline{F'}(\underline{\tilde{t}'}) \underset{\mathrm{rebin}}{\rightarrow} \underline{F}(\underline{\tilde{t}}) \simeq \underline{\tilde{F}}(\underline{\tilde{t}})
\end{equation}

It can be seen that resampling into $N$ sub-time stamps will increase the 
computation time by a factor of $\sim N$, since typically the MA02 subroutine 
uses the majority of a light-curve fitting algorithm's resources (especially for 
non-linear limb darkening). In the next subsection, we will show that the 
computation times can be decreased by \emph{selective resampling}.

One advantage of resampling is that we can choose to resample in such a way as 
to account for read-out and dead-times, which may be important if the 
instrument's duty cycle is quite poor\footnote{We note that this is not the case 
for \emph{Kepler} which has a duty cycle of 91.4\%.}. The resampling of the 
$i^{\mathrm{th}}$ time stamp into $N$ sub-time stamps with labels 
$j=1,2,...N-1,N$ can be expressed as:

\begin{equation}
t_{i,j} = t_i + \Big(j - \frac{N+1}{2}\Big) \frac{\mathcal{I}}{N}
\end{equation}

The flux of the $i^{\mathrm{th}}$ time stamp is found by rebinning all $N$ flux 
stamps from $j=1$ to $j=N$.

\begin{equation}
\tilde{F}_i = \frac{\sum_{j=1}^N F_{i,j}}{N}
\end{equation}

Thus for the first few values of $N=2$, $N=3$ and $N=4$ we would have:

\begin{align}
\tilde{F}_i(N=2) &= \frac{1}{2} \Big[ F(t_i - \mathcal{I}/4) + F(t_i + \mathcal{I}/4) \Big] \\
\tilde{F}_i(N=3) &= \frac{1}{3} \Big[ F(t_i - \mathcal{I}/3) + F(t_i) + F(t_i + \mathcal{I}/3) \Big] \\
\tilde{F}_i(N=4) &= \frac{1}{4} \Big[ F(t_i - 3\mathcal{I}/8) + F(t_i - \mathcal{I}/8) \nonumber \\
\qquad& + F(t_i + \mathcal{I}/8)+F(t_i + 3\mathcal{I}/8) \Big]
\end{align}

For a trapezoid approximated light-curve, it can be easily shown that the error 
in these expressions, as a function of $N$, is given by:

\begin{equation}
\underset{\mathrm{Resampling}}{\sigma_{\tilde{F}}} = \frac{\delta}{\tau} \frac{\mathcal{I}}{8 N^2}
\end{equation}

Therefore, the resampling method yields greater accuracy than the composite 
Simpson's method. In \citet{kipbak10}, both the resampling and Simpson's 
composite rule were employed in completely independent analyses and the obtained 
results were consistent. Therefore, whilst we are free to use either method 
discussed here, the most efficient approach out of the two is resampling.

\citet{gil10} reported that they used a method for fitting the long-cadence 
light-curve of TrES-2b which we interpret to be equivalent to the resampling 
method. The authors split the LC intervals into 30 contributing sub-intervals 
corresponding to the SC cadence i.e. $N=30$. For the reported LC r.m.s. noise of 
$66$ppm and the system parameters of TrES-2b taken from \citet{win08}, we 
estimate that using $N=5$ would produce a maximum possible error in the most 
unfavourably phased data point of $59$ppm and thus using $N=30$ is excessive for 
this light-curve. Our equations therefore permit for a reduction in 
computational time of 600\%. Such a saving is highly advantageous in MCMC 
fitting, which is inherently expensive on the CPU. For example, \citet{kipbak10} 
found that the typical time to globally fit 10-14 light-curves for each planet 
was around 1-2 weeks on modern CPUs. With the accumulation of 3-4 years of 
transits, CPU efficiency will become increasingly important.

\subsection{Selective Resampling}

Resampling time stamps which satisfy $|t_i - t_C| > (t_T + \mathcal{I})/2$ and 
$|t_i - s_C| > (s_T + \mathcal{I})/2$ is unnecessary since $\tilde{F}_i = F_i = 
1$ in such cases (assuming we have folded multiple transits about the orbital 
period).  Note that we define $s_C$ as the mid-time of the secondary eclipse and 
$s_T$ as the duration from contact point I to IV of the secondary eclipse. We
label this method of optimization as \emph{selective resampling}.

Since $(t_T + s_T)/P \sim 2/(\pi a_R)$, this can reduce the number of time 
stamps which require resampling by an order of magnitude for continuous staring 
telescopes like \emph{Kepler} and \emph{CoRoT}. It should be noted that 
selective resampling will not be possible if the light-curve model includes 
phase variations of the planet e.g. HAT-P-7b, \citet{bor09}.

\section{Applied Example}
\subsection{Target selection}

The use of numerical integration techniques minimizes the error in the retrieved 
transit parameters from the LC data, but there is a fundamental loss of 
information which will mean a certain amount of error is unavoidable. A way to 
demonstrate both this point and the implementation of our numerical techniques 
is to provide an example analysis using the \emph{Kepler} data.

\emph{Kepler} discovers planets in LC mode and characterizes those most 
interesting planets using the SC mode. When \emph{Kepler} began observations, 
three transiting planets were already known to reside in the field-of-view and 
thus \emph{Kepler} was able to observe these objects in SC mode immediately. 
These planets are TrES-2b, HAT-P-7b and HAT-P-11b. Out of these three, only 
TrES-2b's SC data is publicly avaialble and thus will be the subject of our 
analysis in this section.

\subsection{Data handling}

The quarter 0 (Q0) and quarter 1 (Q1) LC and SC photometry was made available as 
part of `Data Release 5' from the \emph{Kepler Mission}. The data includes 18 
transits of TrES-2b with no interruptions during the transit events. We make use 
of the corrected photometry in both cases and details on the various corrective 
procedures can be found in the DR5 handbook. The photometry is normalized by 
dividing each time series by the median of the fluxes and then checks were made 
for any long-term behaviour. We found no significant trends in either time 
series or correlations with the centroid positions.

Preliminary fits are implemented to calculate the residuals and therefore 
identify outlier points. This process is repeated on both sets and points 
deviating by more than 3.5 standard deviations are removed.

\subsection{Lightcurve fitting}

Fits are performed following the same approach as described for method A in 
\citet{kipbak10}. We use a Metropolis-Hastings Markov Chain Monte Carlo 
algorithm with 125,000 trials burning out the first 20\%. In total we fit for 9 
free parameters \{$t_C$,$p^2$,$\Upsilon/R_*$,$b^2$,OOT,$P$,$u_1$,$u_2$\}, where 
$t_C$ is the mid-transit time of the first transit, $p^2$ is the ratio-of-radii 
squared, $\Upsilon/R_*$ is the reciprocal of the half-duration (see 
\citealt{kip10} for details on this parameter), $b^2$ is the square of the 
impact parameter, OOT is the baseline flux, $P$ is the orbital period and $u_1$ 
\& $u_2$ are the quadratic limb darkening coefficients. We use the \citet{man02} 
algorithm to generate limb darkened light-curves and note that because TrES-2b 
is relatively bright for the Kepler field at $V=11.4$, fitting for the limb 
darkening is viable. We also point out that we assume a circular orbit for the 
planet for the purposes of simply comparing the two integration modes. A more 
detailed analysis, including a re-analysis of the RV data and incorporating 
eccentricity, is presented in \citet{kipbak10b}.

We account for the blending of the nearby companion TrES-2/C using the z' 
magnitude differences given in \citet{dae09}. This is achieved by including a 
blending factor $B$ and following the corrective procedure outlined in 
\citet{kiptin10}.

For the SC data, we apply no numerical integration since the cadence is short at 
58.8\,s. For the LC data, we use $N=5$ with the resampling method, as calculated 
in \S\ref{sub:resampling}.

\subsection{Results}

\begin{figure}
\begin{center}
\includegraphics[width=8.4 cm]{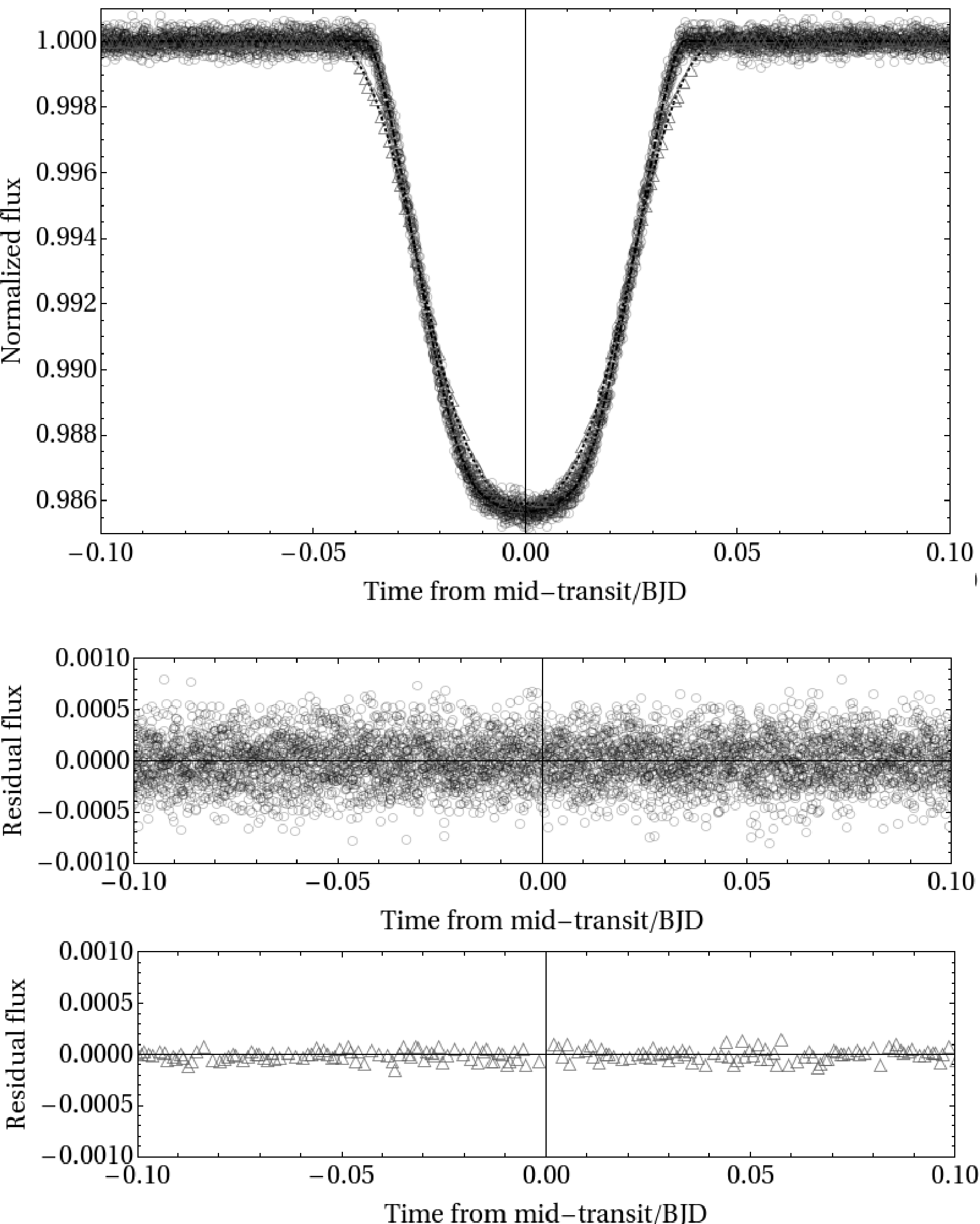}
\caption{\emph{Comparison of the short-cadence data (circles) with overlaid 
best-fit (dashed line) and the long-cadence data (triangles) with overlaid 
best-fit (dotted line). The smearing effect of the long integration times is 
clearly visible, but the retrieved light-curve parameters are consistent using 
the numerical integration techniques outlined in this work.}}
\label{fig:fig3}
\end{center}
\end{figure}

In Table 1, we show the results of our fits for the SC and LC data, visible in 
Figure~\ref{fig:fig3}. We find that comparing parameters from the SC and LC 
modes leads to differences of less than 1\,$\sigma$ in all cases. Therefore, the 
two fits are consistent. We do note that the LC fit led to very poorly 
constrained limb darkening coefficients and thus fitting for limb darkening 
coefficients may not viable for fainter targets in LC mode. This difficulty 
likely stems from the fact limb darkening is suppressed as a consequence of the 
light-curve smearing.

Perhaps not surprisingly, the loss of information caused by binning the data 
leads to larger errors on the parameters in the LC mode. It would therefore seem 
that numerical integration techniques are able to reproduce satisfactory values 
for the light-curve parameters but inevitably lead to larger parameter 
uncertainties.

\begin{table*}
\caption{\emph{Results from global fits of TrES-2b using the short-cadence (SC) 
and long-cadence (SC) data. Orbit is assumed to be circular in all fits. Quoted 
values are medians of MCMC trials with errors given by 1-$\sigma$ quartiles. 
* = fixed parameter; $\dagger$ = parameter was floated but not fitted.}} 
\centering 
\begin{tabular}{c c c c} 
\hline\hline 
Parameter & Short-Cadence & Long-Cadence \\ [0.5ex] 
\hline
\emph{Fitted params.} & & \\
\hline 
$P/$days & $2.4706123_{-0.0000025}^{+0.0000025}$ & $2.4706136_{-0.0000031}^{+0.0000031}$ \\
$t_C$ (BJD-2,454,000) & $955.762539_{-0.000025}^{+0.000025}$ & $955.762530_{-0.000031}^{+0.000031}$ \\
$T_{1,4}$/s & $6442.7_{-14.5}^{+14.8}$ & $6445.6_{-42.6}^{+53.3}$ \\
$T_{1.5,3.5}$/s & $4629.4_{-18.1}^{+19.4}$ & $4660.1_{-77.0}^{+94.0}$ \\ 
$(T_{1,2} \simeq T_{3,4})$/s & $2244.0_{-20.3}^{+21.7}$ & $2187.2_{-114.1}^{+110.6}$ \\
$(R_P/R_*)^2$/\% & $1.632_{-0.032}^{+0.035}$ & $1.580_{-0.122}^{+0.069}$ \\
$b^2$ & $0.7088_{-0.0040}^{+0.0041}$ & $0.707_{-0.020}^{+0.017}$ \\
$(\Upsilon/R_*)$/days$^{-1}$ & $37.33_{-0.16}^{+0.15}$ & $37.08_{-0.73}^{+0.62}$ \\
$B$ & $1.04246 \pm 0.00023$ $\dagger$ & $1.04246 \pm 0.00023$ $\dagger$ \\
\hline
\emph{Limb darkening} & & \\
\hline
$u_1$ & $0.38_{-0.22}^{+0.21}$ & $0.14_{-1.08}^{+0.46}$ \\
$u_2$ & $0.20_{-0.23}^{+0.23}$ & $0.46_{-0.49}^{+1.02}$ \\
\hline
\emph{Model indep. params.} & & \\
\hline
$R_P/R_*$ & $0.1278_{-0.0013}^{+0.0014}$ & $0.1257_{-0.0049}^{+0.0027}$ \\
$a/R_*$ & $7.969_{-0.054}^{+0.056}$ & $7.95_{-0.19}^{+0.15}$ \\
$b$ & $0.8419_{-0.0024}^{+0.0024}$ & $0.841_{-0.012}^{+0.010}$ \\
$i$/$^{\circ}$ & $83.936_{-0.057}^{+0.057}$ & $83.93_{-0.21}^{+0.18}$ \\
\hline\hline 
\end{tabular}
\label{tab:global} 
\end{table*}

\section{Conclusions}

We have explored how long-cadence data, with particular focus on \emph{Kepler}, 
causes severe systematic errors in the retrieved physical parameters, unless 
accounted for.  The effect is valid for any finite exposure time but increases 
with longer cadences. Long-cadence data smears out the light-curve morphology, 
which acts to stretch out the ingress/egress duration and suppress limb 
darkening in the light-curve trough.  These two effects act to increase and 
decrease the retrieved impact parameter respectively. Critically, 
overestimating the impact parameter is shown to lead to severe underestimations 
of the stellar density which could lead to planetary candidates being rejected 
on the basis of being unphysical.
 
Numerical integration techniques permit for improved modeling of the transit 
light-curve. We discuss two particular methods, the composite Simpson's method 
and resampling. We provide expressions for estimating the errors of these 
techniques and find that both methods produce an error which scales as $N^{-2}$ 
where $N$ is the numerical resolution of the techniques. Out of these two 
discussed methods, the resampling approach yields a greater efficiency. 
Incorporation of the effects of the finite-integration times for a real 
light-curve is provided by \emph{Kepler} photometry of TrES-2b, where the two 
integration modes are shown to lead to consistent transit parameters using the 
numerical integration technique.

\section*{Acknowledgments}

We thank the referee Ron Gilliland for suggesting the comparison of the SC and 
LC data of TrES-2b, and other useful comments. DMK is supported by STFC, 
University College London, HATNet and the Harvard Smithsonian Center for 
Astrophysics.  Special thanks to Gaspar Bakos for technical discussions on this 
subject.  Thanks to G. Tinetti, J. P. Beaulieu, M. Swain, I. Waldmann and P. 
Deroo for useful comments in preparing this manuscript.

\bsp

\label{lastpage}


\begin{thebibliography}{99}
\bibitem[\protect\citeauthoryear{Agol et al.}{2005}]{ago05} Agol, E., Steffen, J., Sari R. \& Clarkson W. 2005, MNRAS, 359, 567
\bibitem[\protect\citeauthoryear{Borucki et al.}{2009}]{bor09} Borucki, W. J. et al., Science, 325, 709
\bibitem[\protect\citeauthoryear{Borucki et al.}{2010}]{bor10} Borucki, W. J. et al., Science, 327, 977
\bibitem[\protect\citeauthoryear{Carter et al.}{2008}]{car08} Carter, J. A., Yee, J. C., Eastman, J., Gaudi, B. S. \& Winn, J. N., 2008, ApJ, 689, 499
\bibitem[\protect\citeauthoryear{Carter \& Winn}{2010}]{car10} Carter, J. A. \& Winn, J. N., 2010, ApJ, 709, 1219
\bibitem[\protect\citeauthoryear{Charbonneau et al.}{2000}]{cha00} Charbonneau, D., Brown, T. M., Latham, D. W., \& Mayor, M.  2000, ApJL, 529, L45
\bibitem[\protect\citeauthoryear{Daemgen et al.}{2009}]{dae09} Daemgen, S., Hormuth, F., Brandner, W., Bergfors, C., Janson, M., Hippler, S. \& Henning, T., 2009, A\&A, 498, 567
\bibitem[\protect\citeauthoryear{Deming et al.}{2006}]{dem06} Deming, D., Harrington, J., Seager, S. \& Richardson, L. J.,  2006, ApJ, 644, 560
\bibitem[\protect\citeauthoryear{O'Donovan et al.}{2007}]{don07} O'Donovan, F. T. et al., 2007, ApJ, 651, 61
\bibitem[\protect\citeauthoryear{O'Donovan et al.}{2010}]{don10} O'Donovan, F. T., Charbonneau, D., Harrington, J., Madhusudhan, N., Seager, S., Deming, D., Knutson, H. A., 2010, ApJ, 710, 1551
\bibitem[\protect\citeauthoryear{Gilliland et al.}{2010}]{gil10} Gilliland, R. L. et al., 2010, ApJL, 713, 160
\bibitem[\protect\citeauthoryear{Henry et al.}{2000}]{hen00} Henry, G. W., Marcy, G. W., Butler, R. P. \& Vogt, S. S., 2000, ApJ, 529, L41
\bibitem[\protect\citeauthoryear{Holman \& Murray}{2005}]{hol05} Holman, M. J. \& Murray, N. W., 2005, Science, 307, 1288
\bibitem[\protect\citeauthoryear{Kipping}{2008}]{kip08} Kipping, D. M., 2008, MNRAS, 389, 1383, (K08)
\bibitem[\protect\citeauthoryear{Kipping}{2009a}]{kip09a} Kipping, D. M., 2009a, MNRAS, 392, 181
\bibitem[\protect\citeauthoryear{Kipping}{2009b}]{kip09b} Kipping, D. M., 2009b, MNRAS, 396, 1797
\bibitem[\protect\citeauthoryear{Kipping}{2010}]{kip10} Kipping, D. M., 2010, MNRAS, 407, 301
\bibitem[\protect\citeauthoryear{Kipping \& Bakos}{2011a}]{kipbak10} Kipping, D. M. \& Bakos, G. A., 2011a, ApJ, 730, 50
\bibitem[\protect\citeauthoryear{Kipping \& Bakos}{2011b}]{kipbak10b} Kipping, D. M. \& Bakos, G. A., 2011b, ApJ, 733, 36
\bibitem[\protect\citeauthoryear{Kipping \& Tinetti}{2010}]{kiptin10} Kipping, D. M. \& Tinetti, G., 2010, MNRAS, 407, 2589
\bibitem[\protect\citeauthoryear{Koch et al.}{2010}]{koc10} Koch, D. et al., 2010, ApJL, 713, 131
\bibitem[\protect\citeauthoryear{Mandel \& Agol}{2002}]{man02} Mandel, K. \& Agol, E. 2002, ApJ, 580, L171, (MA02)
\bibitem[\protect\citeauthoryear{McKeeman}{1962}]{mck62} McKeeman, W. M., 1962, Communications of the ACM, 5, 604
\bibitem[\protect\citeauthoryear{Seager \& Hui}{2002}]{sea02} Seager, S., \& Hui, L., 2002, ApJ, 572, 540
\bibitem[\protect\citeauthoryear{Seager \& Mall\'{e}n-Ornelas}{2003}]{sea03} Seager, S., \& Mall\'{e}n-Ornelas, G., 2003, ApJ, 585, 1038
\bibitem[\protect\citeauthoryear{Tinetti et al.}{2007}]{tin07} Tinetti, G., Vidal-Madjar, A., Liang, M.-C. et al., 2007, Nature 448, 169
\bibitem[\protect\citeauthoryear{Winn et al.}{2005}]{win05} Winn, J. N., et al. 2005, ApJ, 631, 1215
\bibitem[\protect\citeauthoryear{Winn et al.}{2008}]{win08} Winn, J. N., et al. 2008, ApJ, 682, 1283
\bibitem[\protect\citeauthoryear{Winn}{2010}]{win10} Winn, J. N., 2010, \emph{Transits and Occultations}, EXOPLANETS, University of Arizona Press; ed: S. Seager
\end{thebibliography}
\end{document}